\documentclass[prl,twocolumn,floatfix,preprintnumbers,amsmath,amssymb,superscriptaddress,nofootinbib]{revtex4-1}

\usepackage{graphicx}
\usepackage{dcolumn}
\usepackage{bm}
\usepackage{array}
\usepackage{multirow}
\usepackage{hyperref}
\usepackage{url}
\usepackage{color}
\usepackage{float}
\usepackage[running]{lineno}
\usepackage{notes2bib}
\usepackage[table,xcdraw]{xcolor}

\newcommand{\tbt}{TbTe$_{3}$\,}
\newcommand{\qc}{\textbf{q$_{c}$\,}}
\newcommand{\qa}{\textbf{q$_{a}$\,}}
\newcommand{\qmc}{\textbf{q$_{m1}$\,}}
\newcommand{\qma}{\textbf{q$_{m2}$\,}}
\newcommand{\cqmc}{\textbf{q$_{cm1}$\,}}
\newcommand{\cqmcc}{\textbf{q$_{cm2}$\,}}
\newcommand{\cqmca}{\textbf{q$_{am2}$\,}}

\begin{document}
\title{Strongly coupled charge, orbital and spin order in \tbt}

\author{S.~Chillal}
\email[]{shravani.chillal@helmholtz-berlin.de}
\affiliation{Helmholtz-Zentrum Berlin f\"ur Materialien und Energie, Hahn-Meitner Platz 1, 14109 Berlin, Germany}

\author{ E.~Schierle}
\affiliation{Helmholtz-Zentrum Berlin f\"ur Materialien und Energie, Hahn-Meitner Platz 1, 14109 Berlin, Germany}

\author{ E.~Weschke}
\affiliation{Helmholtz-Zentrum Berlin f\"ur Materialien und Energie, Hahn-Meitner Platz 1, 14109 Berlin, Germany}

\author{ F.~Yokaichiya}
\affiliation{Helmholtz-Zentrum Berlin f\"ur Materialien und Energie, Hahn-Meitner Platz 1, 14109 Berlin, Germany}

\author{ J.-U.~Hoffmann}
\affiliation{Helmholtz-Zentrum Berlin f\"ur Materialien und Energie, Hahn-Meitner Platz 1, 14109 Berlin, Germany}

\author{O.~S.~Volkova}
\affiliation{M.V.~Lomonosov Moscow State University, Leninskie Gory 1, 119991, Moscow, Russia}
\affiliation{Ural Federal University, 620002 Ekaterinburg, Russia}

\author{A.~N.~Vasiliev}
\affiliation{M.V.~Lomonosov Moscow State University, Leninskie Gory 1, 119991, Moscow, Russia}
\affiliation{Ural Federal University, 620002 Ekaterinburg, Russia}
\affiliation{National Research South Ural State University, 454080 Chelyabinsk, Russia}

\author{A.~A.~Sinchenko}
\affiliation{M.V.~Lomonosov Moscow State University, Leninskie Gory 1, 119991, Moscow, Russia}
\affiliation{Kotel'nikov Institute of Radioengineering and Electronics of RAS, Mokhovaya 11-7, 125009 Moscow, Russia}

\author{P.~Lejay}
\affiliation{Universit\'e Grenoble Alpes, CNRS, Grenoble INP, Institut NEEL, 38042 Grenoble, France}

\author{A.~Hadj-Azzem}
\affiliation{Universit\'e Grenoble Alpes, CNRS, Grenoble INP, Institut NEEL, 38042 Grenoble, France}

\author{P.~Monceau}
\affiliation{Universit\'e Grenoble Alpes, CNRS, Grenoble INP, Institut NEEL, 38042 Grenoble, France}

\author{B.~Lake}
\affiliation{Helmholtz-Zentrum Berlin f\"ur Materialien und Energie, Hahn-Meitner Platz 1, 14109 Berlin, Germany}
\affiliation{Institut f\"ur Festk\"orperphysik, Technische Universit\"at Berlin, 10623 Berlin, Germany}

\begin{abstract}
We report a ground state with strongly coupled magnetic and charge density wave orders mediated via orbital ordering in the layered compound \tbt. In addition to the commensurate antiferromagnetic (AFM) and charge density wave (CDW) orders, new magnetic peaks are observed whose propagation vector equals the sum of the AFM and CDW propagation vectors, revealing an intricate and highly entwined relationship. This is especially interesting given that the magnetic and charge orders lie in different layers of the crystal structure where the highly localized magnetic moments of the Tb$^{3+}$ ions are netted in the Tb-Te stacks, while the charge order is formed by the conduction electrons of the adjacent Te-Te layers. Our results, based on neutron diffraction and resonant x-ray scattering reveal that the charge and magnetic subsystems mutually influence each other via the orbital ordering of Tb$^{3+}$ ions.
\end{abstract}

\date{\today}


\maketitle

Strongly correlated electrons systems, which lie in the poorly-understood region between simple metals and insulators, are home to a rich variety of exotic phases such as charge density waves (CDW), complex magnetic orders and unconventional superconductivity (SC). These phases compete, coexist and cooperate as functions of various tuning parameters leading to rich and often unpredictable phase diagrams~\cite{Fradkin2015,Lian2018}. In the presence of magnetic ions the CDW may influence and be influenced by the magnetic orders such as the appearance of stripe order$-$a collective, long-period modulation of spins and charge carriers within the CuO$_{2}$ planes observed in cuprate systems~\cite{Tranquada1995}. It is also closely associated with superconductivity which appears nearby in the phase diagram~\cite{Taillefer2010,Keren2019}. In many cases the superconductivity is unconventional and is possibly mediated via magnetism as proposed for layered transition-metal chalcogenides, pnictides and copper-oxide high-$T_{c}$ superconductors.

Rare earth chalcogenides of type RTe$_{3}$ that host the three collective orders {\it viz.} CDW, magnetism and unconventional superconductivity, are equally fascinating although much less understood~\cite{Brouet2008,Ru2008b,Ru2008,Hamlin2009,Zocco2015}. Even though the magnetic and superconducting/charge constituents are well-separated as in heavy-fermion systems, the RTe$_{3}$ compounds show no evidence for heavy Fermion behavior~\cite{Hamlin2009,Zocco2009,Ying2020}. Furthermore, no strong correlations have been found between the magnetic rare earth layer and the CDW layers~\cite{Ru2008b,Ru2008,Pfuner2012}. On the other hand, their pressure-dependent phase diagram largely replicates that of the cuprates~\cite{Hamlin2009,Zocco2015} casting the RTe$_{3}$ as ideal systems to understand the interplay of multiple degrees of freedom. In the following, we explore \tbt which is a prominent example of these layered compounds and show that charge and magnetic orders in this material are highly entwined. Furthermore, our investigation, reveals an ubiquitous fourth electronic order involving the Tb-$4f$ orbitals, which plays the crucial role of mediating the order parameters of this system. While the orbital order manifesting as electronic nematic order~\cite{Fradkin2010} has been associated with the rotational symmetry breaking of the $3d$ orbitals in cuprates~\cite{Daou2010,Achkar2016} and Fe-based superconductors~\cite{Chu2010,Liu2019}, its importance in the interplay of these phases is still unclear. Therefore, the role of orbital order in \tbt highlights a new mechanism for the coupling of charge and spin orders compared to the cuprates and heavy fermion superconductors~\cite{Wirth2016,Steglich2016}.

\begin{figure}
\centering
\includegraphics[width=0.45\textwidth]{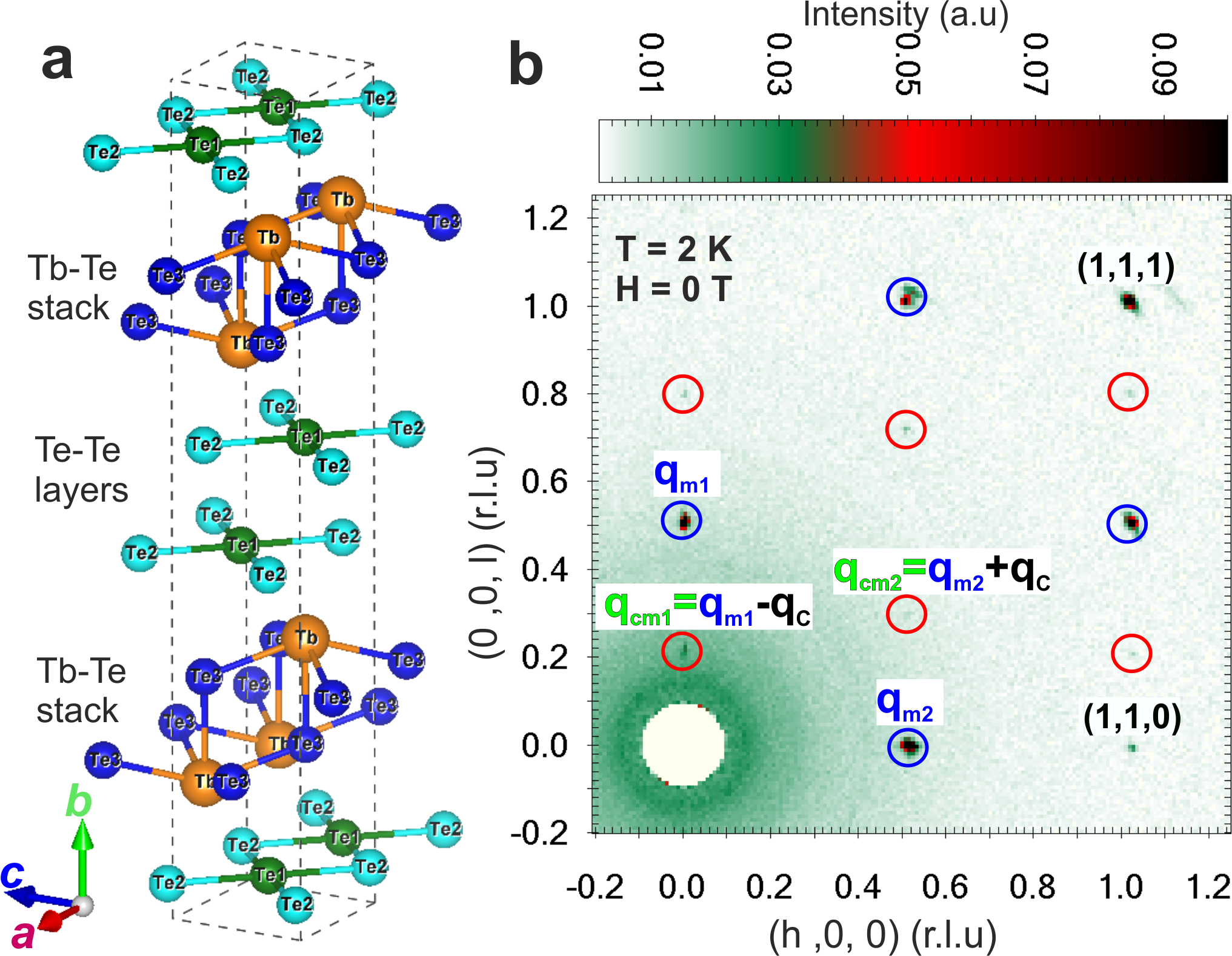}
\caption{\textbf{a}) Unit cell of \tbt where the two-dimensional Te-Te sheets sandwich the Tb-Te stacks. \textbf{b}) Neutron diffraction map for the single crystal of \tbt measured at E2 diffractometer in the $\bm{a}-\bm{c}$ plane at $T=2$~K with intensities as indicated in the colorbar. The intensities are integrated over the out-of-plane $\bm{b}$-axis and reveal CMM peaks with propagation vectors $\cqmc=(0,0,0.21)$, $\cqmcc=(0.5,0.5,0.29)$ (red circles) and AFM peaks $\qmc=(0,0,0.5)$, $\qma=(0.5,0.5,0)$ (blue circles).}
\label{Figure:1}
\end{figure} 

\tbt crystallizes in an orthorhombic structure (spacegroup {\it Cmcm}, lattice parameters $\bm{a}$=4.298\AA, $\bm{b}$=25.33\AA and $\bm{c}$=4.303\AA) as depicted in fig~\ref{Figure:1}a where the quasi two dimensional (2D) nature of the system is evident from the stacking of Te-Te layers and Tb-Te units along the $\bm{b}-$axis. Below $T_{c}=330$~K, a CDW develops along $\bm{c}-$direction with propagation vector $\qc=(0,0,0.296)$ as seen by hard X-ray, electron diffraction, scanning tunneling microscopy as well as  optical conductivity~\cite{DiMasi1995,Fang2007,Ru2008b}. It is directly connected with the nesting of the Fermi surface formed by the Te($5p$) bands of the Te-Te sheets~\cite{Laverock2005,Brouet2008}, with an important role of momentum dependent electron-phonon interactions~\cite{Schmitt2008,Maschek2015}. Additional 2nd and 3rd CDWs are formed at lower temperatures~\cite{Hu2014}, in particular also along the $\bm{a}$-axis with $\qa=(0.32,0,0)$~\cite{Fang2007,Banerjee2013}. 

The magnetic Tb$^{3+}$ ions in Tb-Te layer give rise to three consecutive antiferromagnetic transitions at $T_{N1}\sim6.6$~K, $T_{N2}\sim5.6$~K and $T_{N3}\sim5.4$~K, as seen in heat capacity and resistivity measurements~\cite{Ru2008}. An initial neutron diffraction study~\cite{Pfuner2012} revealed two magnetic propagation vectors $(0,0,0.5)$ and $(0,0,0.21)$ at base temperature. The magnetic structures, however, have not been solved.

In this work, a first overview of the magnetic Bragg peaks was obtained by neutron diffraction. The data were recorded from single crystals of \tbt at $T=2$~K using the E2 Flatcone neutron diffractometer at HZB~\cite{supp}. Figure~\ref{Figure:1}b shows the diffraction map in the $\bm{a}-\bm{c}$ plane revealing several nuclear peaks, commensurate antiferromagnetic (AFM, blue circles) and incommensurate (red circles) magnetic Bragg peaks which also include peaks with an out-of-plane $\bm{b}-$axis component. We find new commensurate and incommensurate magnetic Bragg peaks with propagation vectors $\qma=(0.5,\pm0.5,0)$ and $\cqmcc=(0.5,0.5,0.29)$, respectively, in addition to the previously reported commensurate $\qmc=(0,0,0.5)$ and incommensurate $\cqmc =(0,0,0.21)$ orders~\cite{Pfuner2012}. Further diffraction peaks were found by resonant elastic x-ray scattering (REXS) at the Tb-M$_{5}$ resonance, demonstrating a complex ordering pattern of AFM commensurate and CDW-related incommensurate diffraction peaks. These are summarized in Table.~\ref{tab1}. 

\begin{table}
\scriptsize
\centering
\begin{tabular}{| p{2.2cm} | p{1.9cm} | p{2.15cm} | p{1.95cm}|}
\hline
\bf Type of order &
  \bf Experimental method &
  \bf Ordering wave vector &
  \bf Ordering temperature \\ \hline
  \rowcolor{black!30}
\begin{tabular}[c]{@{}c@{}}Charge density \\ wave (CDW)\end{tabular} &
  \begin{tabular}[c]{@{}c@{}}Hard x-ray\\ diffraction\end{tabular} &
  \begin{tabular}[c]{@{}c@{}}\qc$\approx(0,0,0.29)$\\ \qa$\approx(0.32,0,0)$\end{tabular} &
  \begin{tabular}[c]{@{}c@{}}T$_{c}^{c}=323$~K~\cite{Ru2008b}\\ T$_{c}^{a}=40$~K~\cite{Banerjee2013}\end{tabular} \\ \hline
  \rowcolor{red!30}
\begin{tabular}[c]{@{}c@{}}$^*$CDW-induced \\ orbital (COO)\end{tabular} &
  RXS (Tb-M$_5$) &
  \begin{tabular}[c]{@{}c@{}}  \qc,2\qc \\ \qa \end{tabular} &
  \begin{tabular}[c]{@{}c@{}}T$_c^c\approx323$~K\\ - \end{tabular} \\ \hline
  \rowcolor{blue!30}
\begin{tabular}[c]{@{}c@{}}$^*$Commensurate \\ antiferromagnetic \\(AFM)\end{tabular} &
  \begin{tabular}[c]{@{}c@{}}RXS (Tb-M$_5$)\\ \& Neutron \\ diffraction\end{tabular} &
  \begin{tabular}[c]{@{}c@{}}\qmc$=(0,0,\frac{1}{2})$\\ \\ \qma$=(\frac{1}{2},\frac{1}{2},0)$\end{tabular} &
  \begin{tabular}[c]{@{}c@{}}T$_{N2}=5.7$~K \\ (stable T$<$T$_{N3}$)\\ T$_{N1}=6.7$~K \\ (stable T$<$T$_{N2}$)\end{tabular} \\ \hline
  \rowcolor{green!30}
\begin{tabular}[c]{@{}c@{}}$^*$CDW-modulated \\ AFM (CMM)\end{tabular} &
  \begin{tabular}[c]{@{}c@{}}REXS (Tb-M$_5$)\\ \& Neutron \\ diffraction\end{tabular} &
  \begin{tabular}[c]{@{}c@{}}\cqmc$=\qmc\pm\qc$,\\              $\qmc\pm$2\qc\\ \cqmcc$=\qma\pm\qc$\\ \cqmca$=\qma\pm\qa$\end{tabular} &
  \begin{tabular}[c]{@{}c@{}}T$_{N3}=5.4$~K\\ \\ T$_{N3}=5.4$~K\\ T$_{N3}=5.4$~K\end{tabular} \\ \hline
\end{tabular}
\caption{Summary of the observed CDW, COO, AFM and CMM peaks in \tbt. $^*$ This work.}
\label{tab1}
\end{table}

A closer inspection of Table.~\ref{tab1} reveals that all incommensurate magnetic propagation vectors can be written in terms of the AFM and CDW orders as $\cqmc=\qmc\pm\qc$ and $\cqmcc=\qma\pm\qc$ where $\qc=(0,0,0.29)$, and $\cqmca =\qma\pm\qa$ where $\qa=(0.29,0,0)$ (as observed in the RXS). These CDW-modulated magnetic peaks are termed here CMM. 

\begin{figure}
\centering
\includegraphics[width=0.5\textwidth]{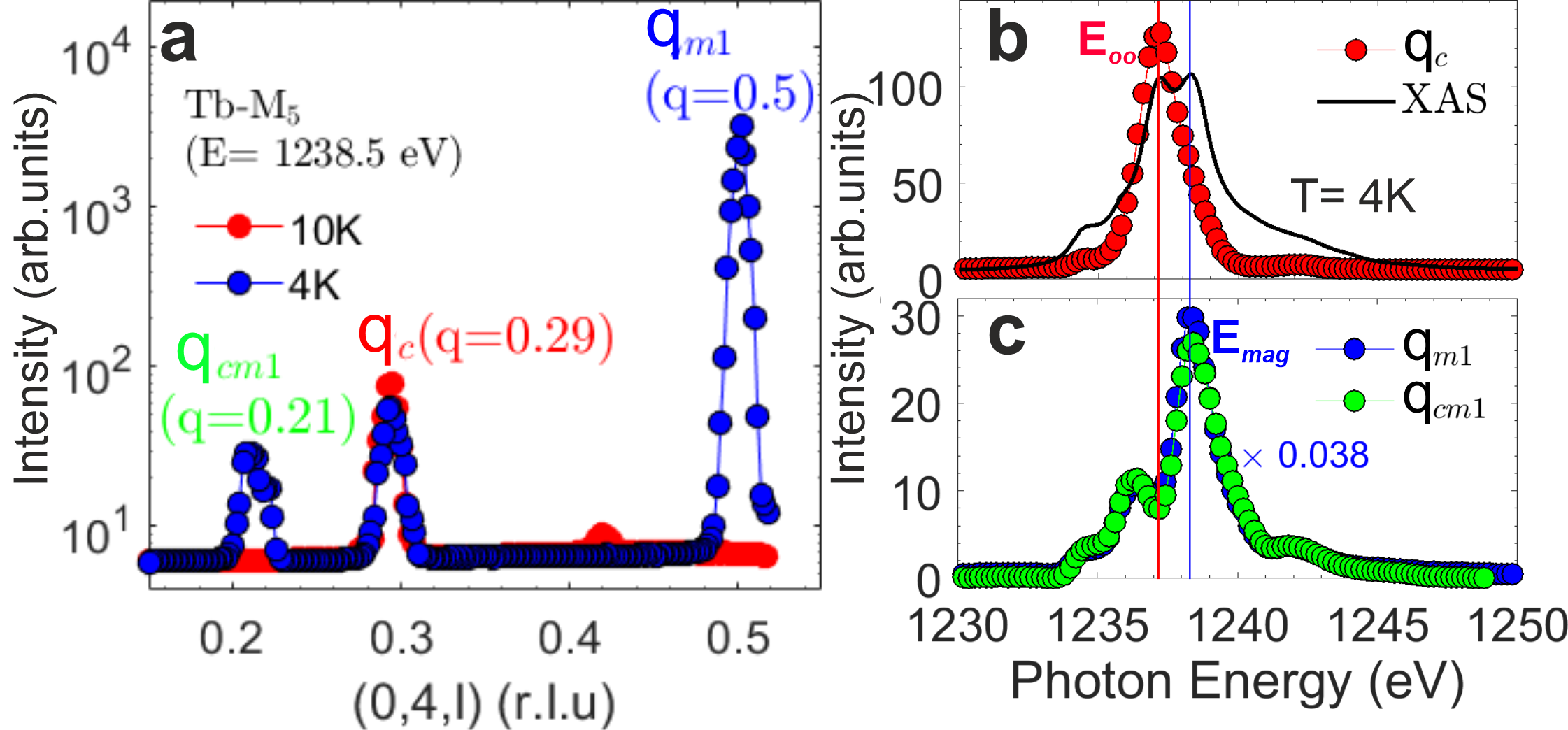}
\caption{Soft x-ray resonant diffraction peaks measured at UE46-PGM1 beamline: \textbf{a}) show the COO$-$\textcolor{red}{\rm $\qc$}, CMM$-$\textcolor{green}{\rm $\cqmc$} and AFM$-$\textcolor{blue}{\rm $\qmc$} peaks along the $\bm{c}-$axis measured at $T=4$~K below $T_{N3}=5.4$~K. The resonance profile, \textbf{b}) of the COO peak at fixed-Q plotted against the Tb-M$_{5}$ absorption edge (black solid line), indicating a maximum at $E_{oo}=1237.3$~eV, and \textbf{c}) of the magnetic peaks featuring a distinctly different multi-peak profile with a maximum at $E_{mag}=1238.5$~eV. All other lines are guides to the eye.}
\label{Figure:2}
\end{figure}

While there is thus clear evidence for coupling of the magnetic ordering in \tbt to the CDW, neutron diffraction is insensitive to charge and orbital modulations. Therefore, resonant x-ray diffraction was used to elucidate the role of these electronic degrees of freedom in the observed coupling. In particular, the Tb-$4f$ states were addressed by tuning the photon energy to the Tb-M$_{5}$ edge ($3d\rightarrow4f$ transition) shown as the x-ray absorption spectra (XAS) in fig.~\ref{Figure:2}b (the solid black line). The experiments were carried out using the XUV and the High-Field diffractometer at the UE-46 beamline of BESSY II at HZB.

As shown in fig.~\ref{Figure:2}a, at resonance we observe not only the AFM peak at \qmc and the CMM peak at \cqmc but in addition a peak at the wave vector transfer of the CDW \qc.  Resonant energy profiles across the Tb-M$_{5}$ resonance were recorded to further characterize these diffraction peaks according to their charge/orbital/magnetic contribution~\cite{Goedkoop1988,Schierle2010,supp}. As shown in fig.~\ref{Figure:2}b and 2c, the AFM peak exhibits a multi-peak structure with a maximum at $E_{mag}=1238.5$~eV while the peak at \qc shows a single peak at $E_{oo}=1237.3$~eV. These energy profiles are consistent with the absorption cross-section calculations of the ferromagnetically ordered Tb$^{3+}$ ions~\cite{Goedkoop1988} and can be identified as magnetic and orbital in origin, respectively. Interestingly, the CMM peak at \cqmc exhibits the same energy profile as the AFM peak at \qmc, showing that it is also predominantly magnetic in nature and originates from a modulation of the Tb$^{3+}$ magnetic moments without any apparent orbital or charge component.  In contrast, the peak observed at \qc is purely of orbital nature. As shown in the SM~\cite{supp}, analogous behavior was also observed for diffraction peaks involving the CDW along $\bm{a}$~\cite{comnt}.

\begin{figure}
\centering
\includegraphics[width=0.5\textwidth]{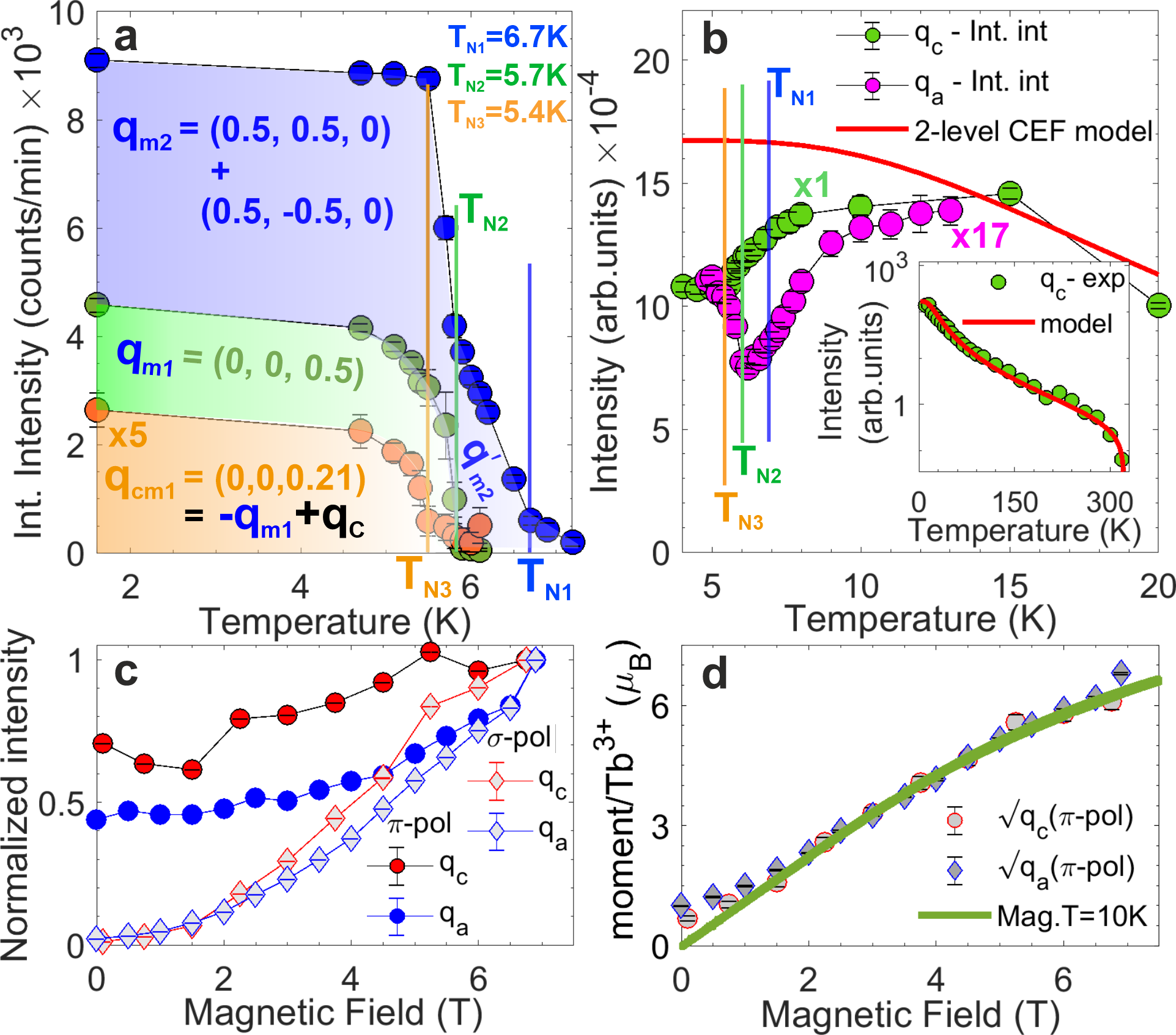} 
\caption{\textbf{a}) Temperature evolution of magnetic peaks \qma, \qmc and \cqmc measured with neutron diffraction indicating three consecutive magnetic transitions. \textbf{b}) the scaled temperature dependence of the COO peaks \qa and \qc from the resonant x-ray scattering manifesting a reduction of intensity in the magnetically ordered state. Inset: the temperature dependence of \qc upon approaching the CDW transition. Solid red lines are the simulated temperature evolution of \qc based on the thermal population of a two-level crystal electric field split (by $4.5$~meV) Tb-4$f$ state~\cite{supp}. \textbf{c}) Magnetic field dependence of the magnetic ($\pi$-pol) and orbital ($\sigma$-pol) contributions to the COO peaks \qa and \qc measured in the paramagnetic state at $10$~K. All the intensities are normalized to their value at $7$~T. \textbf{d}) Square root of the magnetic ($\pi$-pol) intensities as a function of field arbitrarily scaled to match the magnetization of single crystal of \tbt along $\bm{b-}$axis measured at $10$~K. }
\label{Figure:3}
\end{figure}

The intensity of the orbital peak \qc (and \qa) increases exponentially with decreasing temperature and reaches constant (saturated) intensity only at very low temperatures (see inset of figure~\ref{Figure:3}b). Such a behavior has been observed before by Lee et al. and has been analyzed in terms of the temperature-dependent crystal-field level occupation~\cite{Lee2012}. Hence, the intensity of \qc doesn’t track the CDW order parameter itself but instead reflects the degree of CDW-induced orbital order $-$ in close analogy to peaks related to induced magnetic order whose intensity also originates from Boltzmann statistics driven thermal population of magnetic sublevels split by a poling external or internal field. In the following, the peaks representing the CDW-induced $4f$-orbital order at \qc and \qa are therefore termed COO. Consequently, all diffraction peaks of table.~\ref{tab1} represent order in the Tb-Te layers and belong to three categories: (i) COO peaks reflecting the CDW-induced nematicity, (ii) purely AFM peaks at \qmc and \qma, and (iii) CDW-modulated magnetic peaks CMM.

The existence of the COO peaks already provides evidence for a significant impact of the CDWs on the $4f$ orbitals. However, also $4f$-magnetic order should influence the orbital order pattern and vice-versa. This can be inferred from the temperature dependencies of AFM, COO and from the behavior of COO in an external magnetic field. Figure.~\ref{Figure:3}a summarizes the sequence of magnetic phase transitions of \tbt as seen by neutron diffraction. Upon heating from $2$~K, the commensurate AFM \qmc peaks are stable at $(0,0,0.5)$ up to the transition at $T_{N3}=5.4$~K above which the wavevector deviates from the commensurate value and the peaks weaken until disappearing at $T_{N2}=5.7$~K. The commensurate \qma peaks are stable at $(0.5,\pm0.5,0)$ up to $T_{N2}$ above which the wavevector shifts to the in-plane value \textbf{$q'_{m2}$}$=(0.5,0,0)$ and then disappears above $T_{N1}=6.7$~K (see details in~\cite{supp}). In contrast, the incommensurate CMM peaks \cqmc, \cqmca and \cqmcc can only be observed below the lowest transition $T_{N3}$ showing that they can only be stabilized once the commensurate magnetic orders are established. 

The temperature dependence of the COO peak at high temperatures can be described by assuming a two-level crystal-field scenario with a splitting of $4.5$~meV - a model that already captures all essential features of the temperature dependence~\cite{Lee2012} and is plotted as a guide in fig~\ref{Figure:3}b (red solid curve) representing the expected behavior of the diffraction peak at \qc at low temperatures (inset shows experimental data at higher temperatures). Any strong deviation from this behavior must be attributed to the $4f$ magnetic transitions shown in fig~\ref{Figure:3}a. This is in fact seen for both the COO peaks at \qa and \qc. The intensity at \qa drops sharply on approaching the first transition $T_{N1}$ from high temperature. The intensity stays nearly constant below $T_{N1}$ then grows below $T_{N2}$ and finally saturates below $T_{N3}$. Whereas, the peak at \qc decreases relatively slowly reaching a minimum at $T_{N3}$=5.5~K and stabilizes below this temperature. These pronounced modulations are clear evidence that the interaction between the magnetic and orbital orders is in fact mutual and is present for the different spin configurations in all three magnetic phases.

\begin{figure}
\centering
\includegraphics[width=0.5\textwidth]{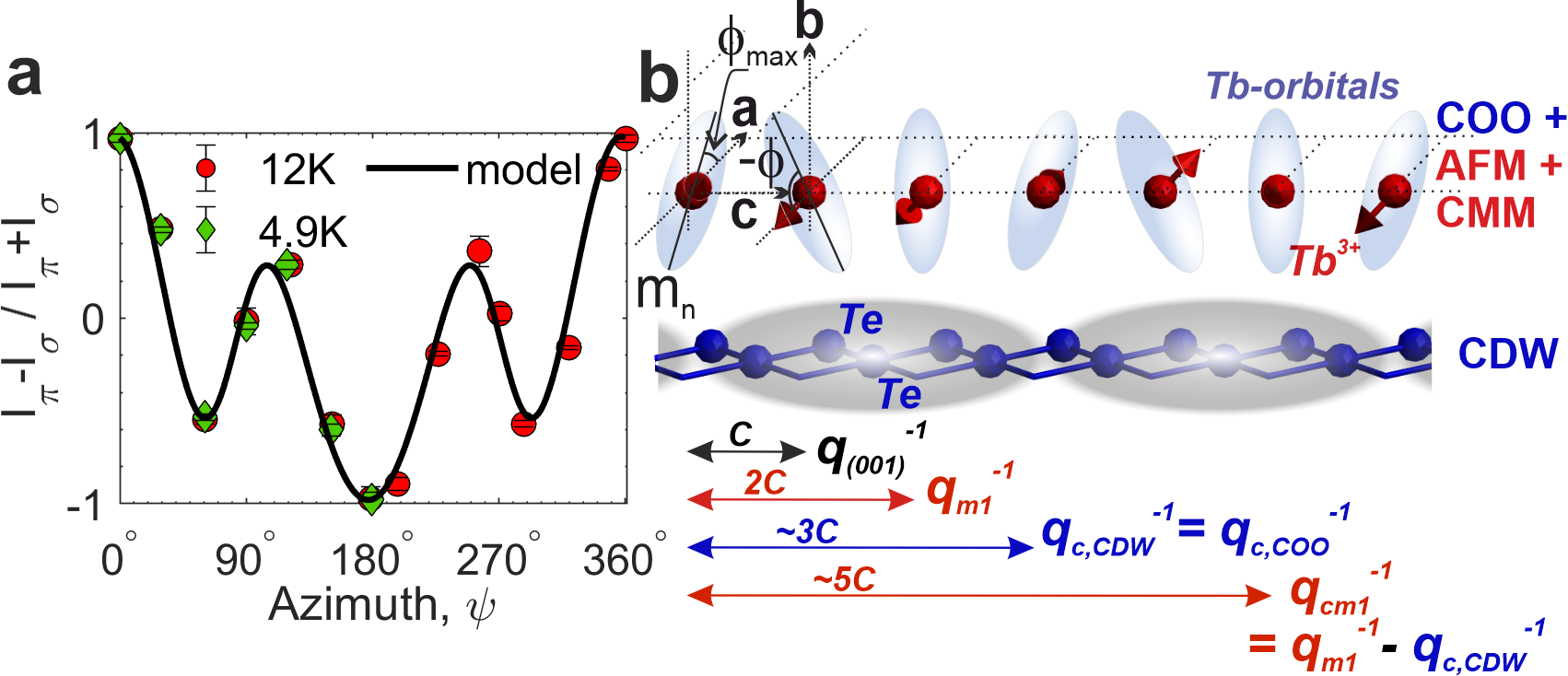}
\caption{\textbf{a}) Azimutal dependence of \qc measured above and below the magnetically ordered state in zero field. Black solid line is the simulation as described in the text and SM~\cite{supp}. \textbf{b}) The schematic of the periodicities of COO (represented by local quantization axis$-m_{i}$ of the Tb-$4f$ orbitals$-$blue ellipsoids) with tilt angle $\phi_i$ about the crystalline $\bm{a}$-axis, AFM and CMM of the Tb$^{3+}$ spins (red arrows) in the Tb-Te layer which follow the CDW (grey clouds) periodicity set in the Te-Te layer of \tbt.}
\label{Figure:4}
\end{figure}

So far, the coupling has been observed as magnetic modulations, where the intrinsic AFM order of the $4f$ system needs to accommodate to the incommensurate wave vector of the CDW. It is therefore instructive to see how ferromagnetic order would interact with the CDW. This is achieved by applying an external magnetic field along the $\bm{b}-$axis to ferromagnetically polarize the Tb$^{3+}$ spins at a temperature $T=10$~K i.e., slightly above $T_{N1}$. The diffraction geometry to study the COO peaks at \qa and \qc is chosen such that orbital and magnetic contributions can be separated: The entire orbital contribution in zero field is seen only with vertically polarized ($\sigma$-pol) incident light, while any induced magnetic contribution is observed in the horizontally polarized ($\pi$-pol) channel. The evolution of COO in the field is shown in fig~\ref{Figure:3}c. We observe a substantial growth of intensity (from zero) in the $\pi$-pol channel with a line shape resembling predominantly that of a pure magnetic contribution~\cite{supp}. We hence observe an induced CMM contribution at the COO peak positions that increases smoothly with applied field in a way as to resemble the magnetization of the system measured at $T=10$~K (fig~\ref{Figure:3}d). This agreement shows that the intensity seen in the $\pi$-pol channel now represents new CDW-modulated { \it ferromagnetic} peaks. With applied field, the orbital contribution to the diffraction peaks at \qa and \qc also increases, as seen in the increase in the $\sigma$-pol channel, while retaining the orbital resonance line shape. This is in stark contrast to the intensity drop that occurs at the spontaneous antiferromagnetic ordering below $T_{N1}$ at zero field. Together, these results suggest that the COO competes with the intrinsic antiferromagnetic order but is enhanced by external-field induced ferromagnetism. 

A complete scenario of the coupling between the CDW in the Te-Te layers and the magnetic/orbital order in the Tb-Te layers can be derived from the azimuthal dependence of the COO peaks: fig.~\ref{Figure:4}a shows the normalized differences of the \qc -COO peak intensities for $\sigma-$ and $\pi-$pol of the incident light upon rotation of the sample about \qc in zero applied magnetic field above $T_{N1}$ at $T=12$~K and in the magnetically ordered phase. This azimuthal-angle dependent map of \qc-related x-ray linear dichroism provides information about the spatial symmetry of the orbital order represented by locally modulated quantization axes. The data of fig.~\ref{Figure:4}a can be very well described by an incommensurate modulation of a tilt angle $\phi_i$ of the local quantization axis $m_{i}$ at site $i$ (black solid line on blue ellipsoids) about the crystalline $\bm{a}-$axis in the $\bm{ab}-$plane with a maximum amplitude $\phi_{max}$ as represented by the (see~\cite{supp} for further details). Using normalized intensities here reduces sample-position and beam-footprint dependent artificial variations of absolute intensities with azimuth and in addition intrinsically separates symmetry-changing effects from a mere global scaling of the individual scattering-channel intensities due to changes of $\phi_{max}$. It turns out, that the magnetic order - while having no impact on the symmetry of the orbital azimuthal dependence (fig.~\ref{Figure:4}a) - influences the intensity of the COO diffraction peaks ( fig.~\ref{Figure:3}), i.e., it changes $\phi_{max}$.  

The azimuthal dependence of the COO order, in combination with the observation of AFM and CMM peaks provides a picture of the coupling of the CDW to the magnetic system, as shown in fig.~\ref{Figure:4}b: The periodic modulation of the conduction electrons forming the CDW in the Te-Te layers induces periodic tilts of the $4f$ orbitals in the Tb-Te layers. Due to an Ising-type anisotropy and the strong $4f$ spin-orbit coupling, the $4f$-spins align antiferromagnetically along the $\bm{a}-$direction giving rise to AFM peaks while simultaneously following these periodic tilts and thus generating an additional magnetic modulation along the $\bm{b}-$direction that appears as the CMM peaks. Within this model, the observed magnetism-related changes of the COO intensities are caused by a reduction of $\phi_{max}$ from $\sim7^{\circ}$ at $12$~K to $\sim5^{\circ}$ below $T_{N3}-$ corresponding to an overall energy reduction of $\sim1\%$ of the exchange energy. In contrast, $\phi_{max}$ increases when an external magnetic field along the $\bm{b}-$axis forces the spins and in turn the orbital moments along this direction. The results summarized in fig.~\ref{Figure:4}b rule out other possible scenarios such as magnetic order-induced local Zeeman splitting that modifies the crystal-field scheme, as this would be incompatible with the observed low-temperature variation of COO intensities which show no change of the azimuthal dependence. We would like to point out that the orbital modulations shown in fig.~\ref{Figure:4}b represent a type of nematic order, which is induced by the presence of the CDW in the adjacent layers. As to the origin of this coupling, we may only speculate about the role of the hybridization between Tb-$4f$, Tb-$5d$ and Te-$5p$ states including substantial charge transfer or a spatially modulated electrostatic field or lattice distortion. Although this detail is not clearly accessible in this work, such a hybridization-related CDW-induced lattice distortion has been indeed identified in IrTe$_{2}$~\cite{Takubo2014}.

Given the evidence for a coupling of the CDW to the orbital ordering pattern, the magnetism-induced changes of the latter observed here must create a feedback on the CDW itself. Although, it would be intriguing to study this impact of magnetic order on the CDW directly by, e.g., non-resonant XRD, given the very different involved energy scales, it may turn out that the modification of the CDW in the Te-Te layer is only marginal at ambient pressures. However, such a mutual coupling would reveal the impact of magnetic order on the CDWs necessary to explain the unusual behavior of H$_{C2}$ through the series of ReTe$_3$’s in the high-pressure superconducting phase that occurs after suppression of the CDW ordering temperature to 0~K~\cite{Zocco2015}. Furthermore, this mutual coupling mechanism could also explain the role of spin fluctuations on the observed magnetoresistance in the ReTe$_{3}$ compounds~\cite{Pariari2019,Lei2020,Ying2020} that suggest an inverse relation of the unusually large carrier mobility of the conduction electrons with the moment of the constituting rare-earth ion. Therefore, a complete picture of coupled order parameters in \tbt that includes unconventional superconductivity will strongly benefit from hard x-ray, resonant soft x-ray and neutron scattering experiments performed under hydrostatic pressure.

In summary we find a mechanism demonstrated by the tri-Tellurides whereby a Fermi-surface-nesting related CDW in one layer is able to introduce other orders such as orbital, nematic, magnetic, and lattice-based patterns with the same periodicity in well-separated, adjacent layers. Therefore, the robust coupling mechanism observed in \tbt could point new routes for engineering novel functionality in layered materials and heterostructures. This is particularly useful if the CDW is tunable by an external field and is connected with other useful properties of the system, eg, high carrier mobility, magnetoresistance, insulating behavior or superconductivity, thus allowing them to be controlled.


\begin{acknowledgments}
S. C. thanks K. Prokes for the helpful insights regarding the heavy-fermion physics. This work has been partially supported by the Ministry of Education and Science of the Russian Federation, contracts 02.A03.21.0006 and 02.A03.21.0011. We also acknowledge the support of Russian Foundation for Basic Research through Grants 17-52-150007 and 17-02-00211.
\end{acknowledgments}

\bibliographystyle{apsrev4-1}
%
%

\end{document}